\documentclass[12pt]{article}

\usepackage{graphicx}
\usepackage{dcolumn}
\usepackage{bm}
\usepackage[a4paper,top=3.5cm,bottom=3.2cm,left=2.3cm,right=2.3cm,bindingoffset=5mm]{geometry}
\usepackage{amsfonts}
\usepackage{latexsym}
\usepackage{amsmath}
\usepackage{amssymb}
\usepackage[english]{babel}
\usepackage[latin1]{inputenc}
\usepackage[T1]{fontenc}
\usepackage{slashed}
\usepackage{float}
\usepackage{subfigure}
\usepackage{color}

    \newcommand{\be}{\begin{equation}}
    \newcommand{\ee}{\end{equation}}
    \newcommand{\ba}{\begin{eqnarray}}
    \newcommand{\ea}{\end{eqnarray}}
    \newcommand{\eite}{\end{itemize}}
    \newcommand{\bite}{\begin{itemize}}

\def\ltap{\ \raisebox{-.4ex}{\rlap{$\sim$}} \raisebox{.4ex}{$<$}\ }
\def\gtap{\ \raisebox{-.4ex}{\rlap{$\sim$}} \raisebox{.4ex}{$>$}\ }

\newcommand{\eq}{\begin{eqnarray}}
\newcommand{\en}{\end{eqnarray}}
\newcommand{\bea}{\begin{eqnarray}}
\newcommand{\eea}{\end{eqnarray}}

\def\eV{\mbox{eV}}

\begin{document}

\hfill{{\small Ref. SISSA 04/2017/FISI}}

\hfill{{\small Ref. IPMU17--0014}}


\vspace{1.0cm}
\begin{center}
{\bf{\large An Alternative Method of Determining the 
Neutrino Mass Ordering 
in Reactor Neutrino Experiments
}}\\

\vspace{0.4cm}
S.M. Bilenky$\mbox{}^{a,b)}$, F. Capozzi$\mbox{}^{c,d)}$ 
and S. T. Petcov$\mbox{}^{e,f)}$
\footnote{Also at: Institute of Nuclear Research and
Nuclear Energy, Bulgarian Academy of Sciences, 1784 Sofia, Bulgaria}

\vspace{0.1cm}
$\mbox{}^{a)}${\em Joint Institute for Nuclear Research, Dubna, R-141980, 
Russia.\\}

\vspace{0.1cm}
$\mbox{}^{b)}${\em  TRIUMF 4004, Wesbrook Mall, Vancouver BC, 
V6T 2A3 Canada\\}

\vspace{0.1cm}
$\mbox{}^{c)}${\em  Dipartimento di Fisica e Astronomia ``Galileo Galilei'', Universit\`a di Padova, Via F.\ Marzolo 8, I-35131 Padova, Italy.\\}

\vspace{0.1cm}
$\mbox{}^{d)}${\em  Istituto Nazionale di Fisica Nucleare (INFN), Sezione di Padova, Via F.\ Marzolo 8, I-35131 Padova, Italy.\\}

\vspace{0.1cm}
$\mbox{}^{e)}${\em  SISSA/INFN, Via Bonomea 265, 34136 Trieste, Italy.\\}
%

\vspace{0.1cm}
$\mbox{}^{f)}${\em Kavli IPMU (WPI), The University of Tokyo, Kashiwa,
Chiba 277-8583, Japan.\\
}

\end{center}

\begin{abstract}

We discuss a novel alternative method of determining the 
neutrino mass ordering in medium baseline experiments with 
reactor antineutrinos. Results on the potential sensitivity 
of the new method are also presented. 

\end{abstract}

\baselineskip 12pt

\section{Introduction} 

\vspace{-0.2cm}
 In the present article we consider a complementary method 
to the one proposed in \cite{Petcov:2001sy}
of determination of the  neutrino mass ordering, 
i.e., the type of spectrum the neutrino masses obey, 
in medium baseline experiments with reactor antineutrinos 
\cite{Bilenky:2016nff}. 
The neutrino mass ordering, as is well known,  
is one of the fundamental characteristics of the reference
3-neutrino mixing scheme that still remains undetermined 
experimentally at present (see, e.g., \cite{PDG2016}).
Many basic neutrino physics observables which 
are planned to be measured in currently running and/or upcoming 
neutrino experiments, depend critically on 
the neutrino mass ordering. These include 
the CP violation asymmetry in long baseline neutrino 
oscillation experiments \cite{NOvA2016,DUNE2016,T2HKK2016}, 
the effective Majorana mass in neutrinoless double beta decay 
experiments \cite{Pascoli:2002xyz}, 
the sum of neutrino masses in the case of 
hierarchical neutrino mass spectrum, etc.    
Without the knowledge of what is 
the neutrino mass ordering, or 
the spectrum of neutrino masses, 
it is impossible 
to make progress in understanding the mechanism giving 
rise to nonzero neutrino masses and neutrino mixing.
Determining the type of neutrino mass spectrum  is one of the 
principal goals of the program of future research in neutrino physics 
(see, e.g., \cite{PDG2016,DUNE2016,T2HKK2016,JUNO,RENO50,PINGU,ORCA2016,INO}).

   Within the reference 3-neutrino mixing scheme we are going to consider, 
the neutrino mass spectrum  
is known to be of two varieties: with normal ordering (NO) and 
with inverted ordering (IO). The two possible types of neutrino mass 
spectrum are related to the two possible signs 
of the neutrino mass squared difference 
$\Delta m^2_{31(32)} \equiv m^2_{3} - m^2_{1(2)}$, 
which is associated, e.g., with the dominant oscillations 
of the atmospheric muon neutrinos and anti-neutrinos 
(see, e.g., \cite{PDG2016}).
The sign of $\Delta m^2_{31(32)}$ cannot be determined 
from the existing global neutrino oscillation data. 
In a widely used convention of numbering 
the neutrinos with definite mass 
in the two cases of $\Delta m^2_{31(32)} > 0$ and 
$\Delta m^2_{31(32)} < 0$ 
we are going to employ (see, e.g., \cite{PDG2016}), 
the two possible neutrino mass spectra 
are defined as follows.\\ 
{\it i) Spectrum with normal ordering (NO)}: 
\ba
\label{dm2NO}
m_1 < m_2 < m_3\,,~~~
\Delta m^2_{31(32)} >0\,,~\Delta m^2_{21} > 0\,, \\ [0.3cm]
~~~~~~~~~~~~~~~~m_{2(3)} = (m_1^2 + \Delta m^2_{21(31)})^{1\over{2}}\,. 
\label{m23NO}
\ea
%
{\it ii) Spectrum with inverted ordering (IO)}:  
\ba
\label{dm2IO}
m_3 < m_1 < m_2\,,~~~
\Delta m^2_{32(31)}< 0\,,~\Delta m^2_{21} > 0\,,~~~~\\ [0.30cm]
~~~~~~~~~~~~m_{2} = (m_3^2 - \Delta m^2_{32})^{1\over{2}}\,,~ 
m_{1} = (m_3^2 - \Delta m^2_{32} - \Delta m^2_{21})^{1\over{2}}\,. 
\label{m12IO}
\ea
%
In eqs. (\ref{dm2NO}) - (\ref{m12IO}), 
$\Delta m^2_{21} \equiv \Delta m_{\odot}^2$ 
is the neutrino mass squared difference which is responsible   
for the flavour conversion of the solar electron neutrinos 
(see, e.g., \cite{PDG2016}) and we have expressed the two 
heavier neutrino masses in terms of the lightest neutrino mass 
and the two neutrino mass squared differences measured 
in neutrino oscillation and solar neutrino experiments.

  The existing neutrino oscillation data allow to determine 
the values of $\Delta m^2_{21}$ and $|\Delta m^2_{31(32)}|$, 
as well as the values of the three neutrino mixing angles 
$\theta_{12}$, $\theta_{23}$ and  $\theta_{13}$ of 
the Pontecorvo-Maki-Nakagawa-Sakata (PMNS)
neutrino mixing matrix \cite{BPont57,MNS62,BPont67},
$U_{PMNS}\equiv U$,  
with impressively high precision performing global data 
analyses \cite{Capozzi:Nu2016,Esteban:2016qun,Capozzi:2020qhw}.
The best fit values (b.f.v.) and the 3$\sigma$ allowed ranges of 
$\Delta m^2_{21}$, $\sin^2\theta_{12} \equiv s^2_{12}$, $|\Delta m^2_{31(32)}|$
and $\sin^2\theta_{13} \equiv s^2_{13}$, 
which are relevant for, and will be used in, 
our study read \cite{Capozzi:2020qhw}:
\begin{eqnarray}
\label{deltasolvalues}
&(\Delta m^2_{21})_{\rm BF} = 7.34 \times 10^{-5}\ \eV^2\,,~~~~ 
 \Delta m^2_{21} = (6.92 - 7.90) \times 10^{-5} \ \eV^2\,,\\
\label{sinsolvalues}
&(\sin^2 \theta_{12})_{\rm BF} = 0.305~(0.303)\,,~~
 0.265~(0.264) \leq \sin^2 \theta_{12} \leq 0.347~(0.345)\,,
~\\ 
\label{deltaatmvalues}
&|(\Delta m^2_{31(32)})_{\rm BF}| = 2.522~(2.502) \times 10^{-3} \ \eV^2\,,
~~~~~~~~~~~~~~~~~~~~~~~~~~~~~~~\\ 
\label{deltaatmerrors}
&|\Delta m^2_{31(32)}| = (2.426~(2.412) - 2.615~(2.593))\times 10^{-3}\ \eV^2\,,
~~~~~~~~~~~~~~~\\ 
\label{theta13values}
&(\sin^2\theta_{13})_{\rm BF}=0.022~(0.023)\,,~~ 
0.0201~(0.0203)\leq \sin^2\theta_{13}\leq 0.0241~(0.0243)\,,
\end{eqnarray}
%
where the value (the value in brackets) 
corresponds to  $\Delta m^2_{31(32)}>0$ ($\Delta m^2_{31(32)} <0$).
The quoted values of $\sin^2 \theta_{12}$ and 
$\sin^2 \theta_{13}$ are obtained using the standard parametrisation 
of the PMNS matrix (see, e.g., \cite{PDG2016}). 
We have, in general:  $\sin^2 \theta_{12} = |U_{e2}|^2/(1 - |U_{e3}|^2)$, 
$\sin^2 \theta_{13} = |U_{e3}|^2$, where $U_{e2}$ and $U_{e3}$ 
are elements of the first row of the PMNS matrix.
 
  The possibility to determine the neutrino mass ordering 
in experiments with reactor 
neutrinos  was discussed first in \cite{Petcov:2001sy}.
It was based on the observation made in 
\cite{Bilenky:2001jq} that, given the energy $E$ and 
the distance $L$ travelled by the reactor 
$\bar{\nu}_e$,  
the 3-neutrino mixing probabilities of 
$\bar{\nu}_e$ survival in the cases of NO and IO spectra, 
$P^{NO}(\bar{\nu}_e \rightarrow \bar{\nu}_e)$ and 
$P^{IO}(\bar{\nu}_e \rightarrow \bar{\nu}_e)$, 
differ provided $\sin^2 \theta_{12} \neq \cos^2 \theta_{12}$ 
and  $\sin^2\theta_{13}\neq 0$.
From the data on $\sin^2 \theta_{12}$ 
available already in the second half of 2001 
it followed that the first inequality is fulfilled. 
The two different expressions 
of $P^{NO}(\bar{\nu}_e \rightarrow \bar{\nu}_e)$ and 
$P^{IO}(\bar{\nu}_e \rightarrow \bar{\nu}_e)$ 
were used in \cite{Bilenky:2001jq}
to perform a 3-neutrino mixing analysis 
of the data of the CHOOZ reactor neutrino
experiment \cite{Apollonio:1999ae}, in which the first 
significant constraint on $\sin^22\theta_{13}$ was obtained.

  As is well known, $\bar{\nu}_e$ are detected 
in reactor neutrino experiments of interest 
via the inverse $\beta-$decay
reaction  $\bar{\nu}_e + p \rightarrow e^{+} + n$. 
For protons at rest, 
the $\bar{\nu}_e$ energy $E_{\nu}$ is related to the 
$e^{+}$ energy $E_e$ to a good approximation via 
\ba
E_\nu =  E_{e} + (m_n-m_p)\,,
~~E^{\rm th}_\nu = m_e + (m_n - m_p) \cong 1.8~{\rm MeV}\,,
\label{EeEnu}
\ea
%
where $m_e$, $m_n$ and  $m_p$
are the masses of the positron, neutron and proton and 
$E^{\rm th}_\nu$ is the threshold neutrino energy. 

 In \cite{Petcov:2001sy} it was realised that 
for not exceedingly small $\sin^22\theta_{13}$ 
the difference between 
$P^{NO}(\bar{\nu}_e \rightarrow \bar{\nu}_e)$ 
and $P^{IO}(\bar{\nu}_e \rightarrow \bar{\nu}_e)$ 
leads for medium baselines $L$ 
to a noticeable difference between the NO and IO 
spectra of $\bar{\nu}_e$, and thus of $e^{+}$,
measured in reactor neutrino experiments. 
This led to the conclusion  \cite{Petcov:2001sy} 
that a high precision measurement of the 
$\bar{\nu}_e$ (or $e^{+}$) spectrum 
in a medium baseline reactor neutrino experiment
can provide unique information on the type 
of spectrum neutrino masses obey.
In \cite{Petcov:2001sy} most of the 
numerical results were obtained 
for the value of $\Delta m^2_{21} =2.0\times 10^{-4}~{\rm eV^2}$
from the existing in 2001 ``high-LMA'' region of
solutions  of the solar neutrino problem. 
It was found, in particular, that the optimal source-detector 
distance for determining the neutrino mass ordering
(i.e., the sign of $\Delta m^2_{31(32)}$)  
in the discussed experiment for the 
quoted ``high'' LMA value of $\Delta m^2_{21}$
is $L\sim 20$ km. 
For the current best fit value of 
$\Delta m^2_{21} \cong 7.34\times 10^{-5}~{\rm eV^2}$, 
the optimal distance is by a factor 
$2.0\times 10^{-4}/7.34\times 10^{-5} \cong 2.72$ 
greater: $L\sim 54.5$ km.
Already in \cite{Petcov:2001sy} it was realised 
that the determination of the neutrino mass ordering 
in the proposed reactor neutrino experiment 
would be very challenging from experimental point of view 
(see the ``Conclusion''  section in \cite{Petcov:2001sy}).
 
 A more general and detailed analysis 
performed in \cite{Choubey:2003qx} revealed that 
a reactor neutrino experiment with a medium baseline 
tuned to achieve highest sensitivity 
in the determination of the neutrino mass ordering has 
a remarkable physics potential. 
It was found, in particular, 
that for $\sin^2\theta_{13}\gtap 0.02$ 
it would be possible to measure 
in such an experiment also
i) $\sin^2\theta_{12}$, 
ii) $\Delta m^2_{21}$ and 
iii) $|\Delta m^2_{31(32)}|$  
with an exceptionally high 
precision. It was concluded that the 
precision on  $\sin^2\theta_{12}$ 
and $\Delta m^2_{21}$ that can be achieved 
in the discussed experiment 
cannot be reached in any of the other 
currently running or proposed experiments  
in which these  oscillation 
parameters can be measured. 

 Subsequently, the possibility to determine 
the type of spectrum neutrino masses obey, 
discussed in \cite{Petcov:2001sy,Choubey:2003qx}, 
was further investigated by a number of authors
(see, e.g., \cite{Learned:2006wy,Zhan:2008id,Ghoshal:2010wt}).
It was further scrutinised, e.g., in the studies 
\cite{Qian:2012xh,Ghoshal:2012ju,Ciuffoli:2012iz,Ge:2012wj,Li:2013zyd,Capozzi:2013psa,Capozzi:2015bpa,Wang:2016vua,Blennow:2013oma}, 
after the high precision 
measurements  of $\sin^22\theta_{13}$ 
in 2012 in the Daya Bay and RENO experiments 
\cite{DBay2012,RENO2012}, 
which demonstrated that $\sin^2\theta_{13}\cong 0.025$~
\footnote{Evidence for a non-zero $\theta_{13}$
were obtained earlier in T2K \cite{abe11b} and 
MINOS \cite{adamson11c} accelerator neutrino  
and in Double Chooz \cite{DChoozth13}
reactor neutrino experiments.
}.
These studies showed that 
the determination of the neutrino 
mass ordering in the experiment 
proposed in \cite{Petcov:2001sy}
is indeed very challenging.
It requires:
i) an energy resolution 
$\sigma/E_{\rm vis}\ltap 3\%/\sqrt{E_{\rm vis}/\text{MeV}}$, 
where $E_{\rm vis} = E_e + m_e$ is the ``visible energy'', i.e., 
the energy of the photons emitted when the 
positron produced in the reaction 
$\bar{\nu}_e + p \rightarrow e^{+} + n$ 
annihilates with an electron in the detector; 
ii) a relatively small energy scale uncertainty;
iii) a relatively large statistics ($\sim(300 - 1000)$ 
GW$\times$kton$\times$yr);
iv) relatively small systematic errors;
v) subtle optimisations (e.g., of distances to the reactors 
providing the $\bar{\nu}_e$ flux, of the number of bins,
etc.).

 Two reactor neutrino experiments (employing liquid scintillator 
detectors) with medium baseline of 
$L \cong 50$ km, aiming to determine the neutrino mass ordering, 
have been proposed: JUNO (20 kton) \cite{JUNO}, which is 
approved and is under construction,
and RENO50 (18 kton) 
\footnote{For lack of sufficient funding the work on 
RENO50 project was stopped in the second half of 2017.} 
\cite{RENO50}. In addition of the potential to measure 
$\sin^2\theta_{12}$, $\Delta m^2_{21}$ and $|\Delta m^2_{31(32)}|$ 
with remarkably high precision, these experiments can
be used also for detection and studies of geo, 
solar and supernovae neutrinos.

\vspace{-0.6cm}
\section{Determining the Neutrino Mass Ordering in  
Medium Baseline Reactor Neutrino Experiments} 

\vspace{-0.2cm}
There are two possible consistent 
sets of "atmospheric" neutrino mass squared 
differences in the cases of NO and IO spectra
one can use in the analyses of the 
neutrino oscillation data:
\ba
\label{dm31NOdm23IO}
{\rm NO}:~\Delta m^2_{\mbox{atm}} = \Delta m^2_{31} > 0\,;~~ 
{\rm IO}:~\Delta m^2_{\mbox{atm}} = \Delta m^2_{32} < 0\,,~\\[0.25cm]
\nonumber
{\rm or}~~~~~~~~~~~~~~~~~~~~~~~~~~~~~~~~~~~~~~~~~~~~~~~~~~~~~~~~~~~~~~~~~~~~~~\\[0.25cm] 
{\rm NO}:~\Delta m^2_{\mbox{atm}} = \Delta m^2_{32} > 0\,;~~ 
{\rm IO}:~\Delta m^2_{\mbox{atm}} = \Delta m^2_{31} < 0\,.
\label{dm32NOdm13IO}
\ea
%
In what follows we will use 
the set in eq. (\ref{dm31NOdm23IO}).

 The 3-neutrino mixing 
probabilities of $\bar{\nu}_e$ survival 
in the cases of NO and IO neutrino mass spectra of interest,
$P^{NO}(\bar{\nu}_e \rightarrow \bar{\nu}_e)$ 
and $P^{IO}(\bar{\nu}_e \rightarrow \bar{\nu}_e)$, 
have the following form 
\cite{Bilenky:2001jq,Petcov:2001sy,Choubey:2003qx}:
\bea
\lefteqn{P^{NO}({\bar \nu_e}\to{\bar \nu_e})} \nonumber\\
&& =\,~~ 1 - \frac{1}{2}\, \sin^22\theta_{13}\,
\left( 1 - \cos \frac{ \Delta{m}^2_{\mbox{atm}} \, L }{ 2 \, E_{\nu} } \right)
\nonumber \\
&& - \,~~\frac{1}{2} \cos^4\theta_{13}\,\sin ^{2}2\theta_{12} \,
\left( 1 - \cos \frac{ \Delta{m}^2_{\odot} \, L }{ 2 \, E_{\nu} } \right)
\nonumber \\
& & +\,~~ \frac{1}{2}\,\sin^22\theta_{13}\, \sin^{2}\theta_{12}\,
\left(\cos
\left( \frac
{\Delta{m}^2_{\mbox{atm}} \, L }{ 2 \, E_{\nu}} - 
\frac {\Delta{m}^2_{\odot} \, L }{ 2 \,
E_{\nu}}\right)
-\cos \frac {\Delta{m}^2_{\mbox{atm}} \, L }{ 2 \, E_{\nu}} \right)\, ,
\label{PNO}
\eea
%
\bea
\lefteqn{P^{IO}({\bar \nu_e}\to{\bar \nu_e})} \nonumber\\
&& =\,~~ 1 - \frac{1}{2} \, \sin^22\theta_{13}\,
\left( 1 - \cos \frac{ \Delta{m}^2_{\mbox{atm}} \, L }{ 2 \, E_{\nu} } \right)
\nonumber \\
&& - \,~~\frac{1}{ 2} \cos^{4}\theta_{13}\,\sin ^{2}2\theta_{12} \,
\left( 1 - \cos \frac{ \Delta{m}^2_{\odot} \, L }{ 2 \, E_{\nu} } \right)
\nonumber \\
& & + \,~~\frac{1}{2}\,\sin^22\theta_{13}\, \cos^{2}\theta_{12} \,
\left(\cos
\left( \frac
{\Delta{m}^2_{\mbox{atm}} \, L }{ 2 \, E_{\nu}} - 
\frac {\Delta{m}^2_{\odot} \, L }{ 2 \, E_{\nu}}\right)
-\cos \frac {\Delta{m}^2_{\mbox{atm}} \, L }{ 2 \, E_{\nu}} \right)\,. 
\label{PIO}
\eea
%
As it follows from eqs. (\ref{PNO}) and (\ref{PIO}),
the only difference between the expressions for 
$P^{NO}(\bar{\nu}_e \rightarrow \bar{\nu}_e)$ 
and $P^{IO}(\bar{\nu}_e \rightarrow \bar{\nu}_e)$ is 
in the coefficient of the last terms: 
it is $\sin^2\theta_{12}$ in the NO case and 
$\cos^2\theta_{12}$ in the IO case. 
For the current best fit value of 
$\sin^2\theta_{12} = 0.305~(0.303)$ 
quoted in eq. (\ref{sinsolvalues}) 
we have $\cos^2\theta_{12} \cong 0.695~(0.697)$, 
i.e., the coefficient under discussion 
in $P^{IO}(\bar{\nu}_e \rightarrow \bar{\nu}_e)$ is 
approximately by a factor of 2.3 larger 
than  that in $P^{NO}(\bar{\nu}_e \rightarrow \bar{\nu}_e)$.

 In the standard approach of evaluation of sensitivity 
of a medium baseline 
reactor neutrino experiment to the type 
of spectrum neutrino masses obey,
apart of the specific characteristics of the 
detector, the $\bar{\nu}_e$ flux and energy 
spectrum and their uncertainties, etc., 
one uses as input the
data on the oscillation parameters 
$\sin^{2}\theta_{12}$, $\sin^2\theta_{13}$
$\Delta{m}^2_{\mbox{atm}}$ and $\Delta{m}^2_{\odot}$,  
including the errors with which they are determined (see, e.g., 
\cite{Choubey:2003qx,Ghoshal:2012ju,Ge:2012wj,Li:2013zyd,Capozzi:2013psa}). 
With the indicated inputs one performs 
effectively two statistical 
analyses of prospective (simulated) data, 
generated for a chosen ''true'' 
type of neutrino mass spectrum
\footnote{To simulate prospective rector neutrino data one 
has to  choose one of the two neutrino mass orderings (spectra) 
to be the ''true'' one.}: 
first employing expression 
(\ref{PNO}) for $P^{NO}(\bar{\nu}_e \rightarrow \bar{\nu}_e)$
and thus testing statistically the hypothesis of 
NO spectrum, 
and second - using the expression
(\ref{PIO}) for 
$P^{IO}(\bar{\nu}_e \rightarrow \bar{\nu}_e)$
and testing the possibility of IO spectrum.
The results of these analyses are utilised 
to determine the statistical sensitivity 
of the experiment to the neutrino mass ordering.
In this standard approach the data are used 
to determine an observable - the neutrino mass ordering 
or the sign of $\Delta m^2_{31(32)}$ - which is not 
continuous but can assume just two discrete values.  
Extracting information about 
such a discrete observable requires 
somewhat non-standard statistical methods of 
treatment and interpretation of the data 
\cite{Blennow:2013oma}.

 The alternative method of 
determining the neutrino mass ordering we are going 
to discussed next 
is based on the observation \cite{Bilenky:2016nff}
that the two different expressions (\ref{PNO}) and (\ref{PIO}) for 
$P^{NO}(\bar{\nu}_e \rightarrow \bar{\nu}_e)$ 
and $P^{IO}(\bar{\nu}_e \rightarrow \bar{\nu}_e)$
can be written as:
\bea
\lefteqn{P^{(X)}({\bar \nu_e}\to{\bar \nu_e})} \nonumber\\
&& =\,~~ 1 - \frac{1}{2}\, \sin^22\theta_{13}\,
\left( 1 - \cos \frac{ \Delta{m}^2_{\mbox{atm}} \, L }{ 2 \, E_{\nu} } \right)
\nonumber \\
&& - \,~~2\,\cos^4\theta_{13}\, X^2\,(1 - X^2)\,
\left( 1 - \cos \frac{ \Delta{m}^2_{\odot} \, L }{ 2 \, E_{\nu} } \right)
\nonumber \\
& & +\,~~ \frac{1}{2}\,\sin^22\theta_{13}\, X^2\,
\left(\cos
\left( \frac
{\Delta{m}^2_{\mbox{atm}} \, L }{ 2 \, E_{\nu}} - 
\frac {\Delta{m}^2_{\odot} \, L }{ 2 \,
E_{\nu}}\right)
-\cos \frac {\Delta{m}^2_{\mbox{atm}} \, L }{ 2 \, E_{\nu}} \right)\,,
\label{PNOIO}
\eea
%
where 
\bea 
\label{XNO}  
& X^2 = \sin^2\theta_{12}\,,~~~~~~~{\rm NO~spectrum}\,, \\[0.25cm]
& X^2 = \cos^2\theta_{12}\,,~~~~~~~{\rm IO~spectrum}\,.
\label{XIO}
\eea
%
The determination of the neutrino mass ordering is then 
equivalent to the determination of the value of 
the continuous parameter $X^2$ and comparing the result 
with the value of $\sin^2\theta_{12}$, including its uncertainty, 
determined, e.g., in the solar neutrino and in KamLAND experiments.
Given the fact that, according to the current data, 
the best fit values of 
$\sin^2\theta_{12}$ and  $\cos^2\theta_{12}$
differ by a factor of 2.3, 
and that $\sin^2\theta_{12}$ is determined with 
a $1\sigma$ uncertainty of approximately 4.5\%, 
the proposed method of determining the neutrino mass 
ordering seems feasible. Moreover,
since $X^2$ is a continuous parameter, 
one can use standard statistical methods of 
extracting the value of $X^2$ and its respective 
uncertainty from the data.
The values of $\sin^2\theta_{13}$ and 
$\Delta{m}^2_{\odot}$ measured in independent 
experiments can 
\begin{figure}[t] 
\vspace{-0.5cm}
\begin{center}
\subfigure
{\includegraphics[width=7.8cm]{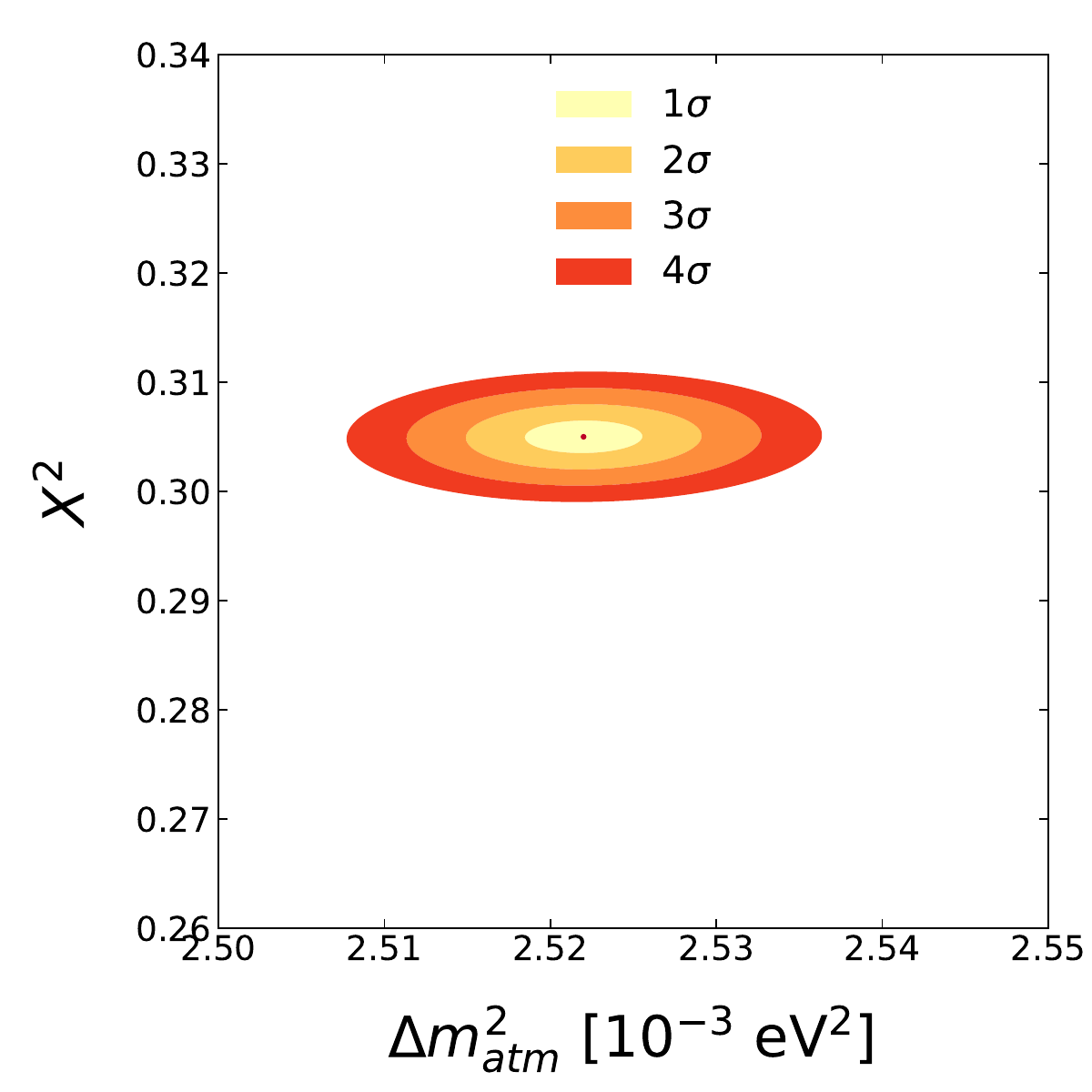}}
{\includegraphics[width=7.8cm]{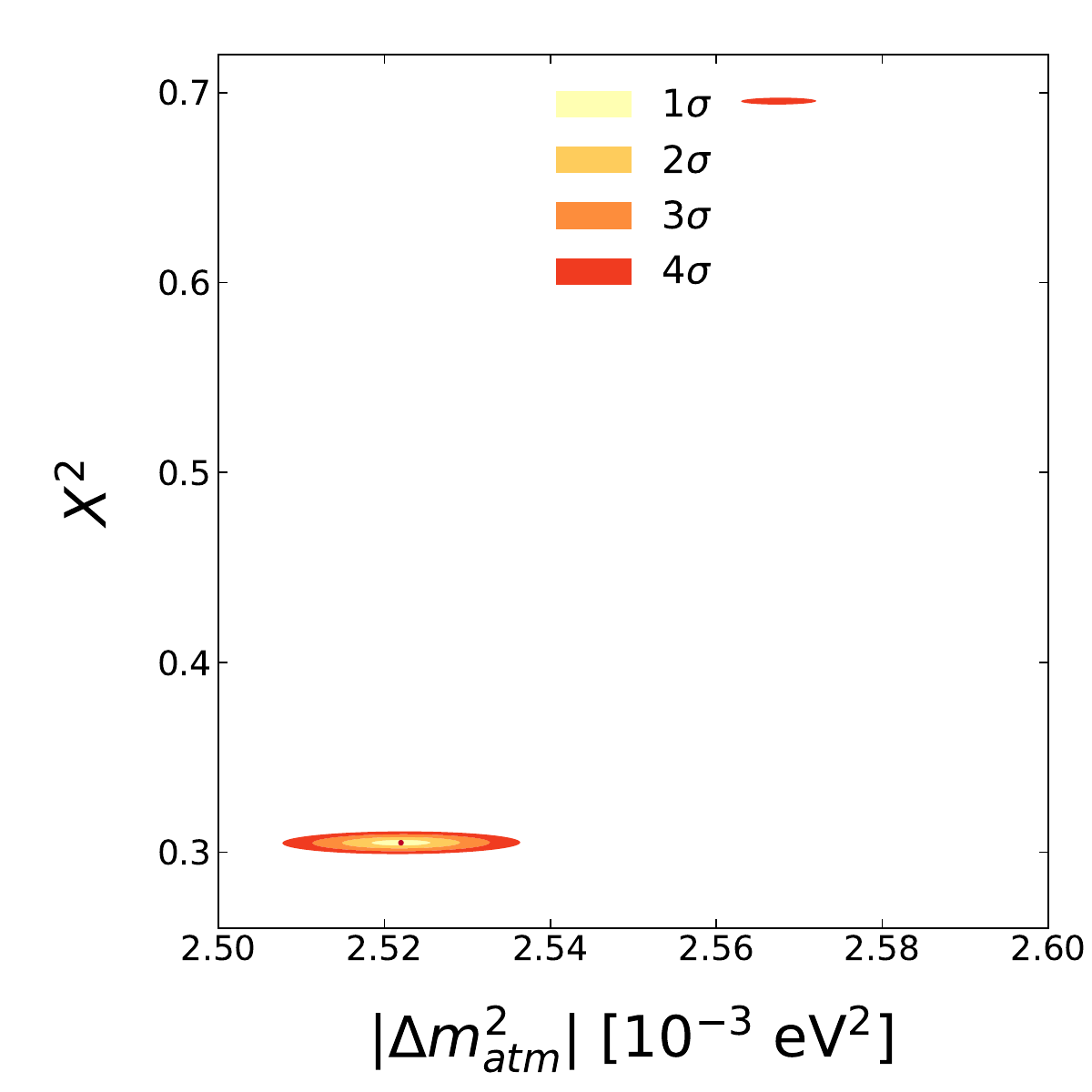}}
\end{center}
\vskip -0.4cm
\caption{The 1$\sigma$, 2$\sigma$, 3$\sigma$ and 
4$\sigma$ C.L. contours in the $\Delta{m}^2_{\mbox{atm}} - X^2$ 
plane obtained in a statistical analysis of prospective 
(simulated) reactor neutrino data from 
JUNO detector. 
The prospective data was generated 
assuming NO neutrino mass 
spectrum and statistics corresponding 
to 3.6$\times 10^3$ GW$\times$kton$\times$yr. 
The second minimum seen in the right panel 
at approximately 4$\sigma$ C.L. 
at $X^2 = 0.695$ corresponds to 
IO neutrino mass spectrum.
See text for further details.
}
\label{Fig1}
\vspace{-0.1cm}
\end{figure}
%
\begin{figure}[t] 
\vspace{-0.5cm}
\begin{center}
\subfigure
{\includegraphics[width=7.8cm]{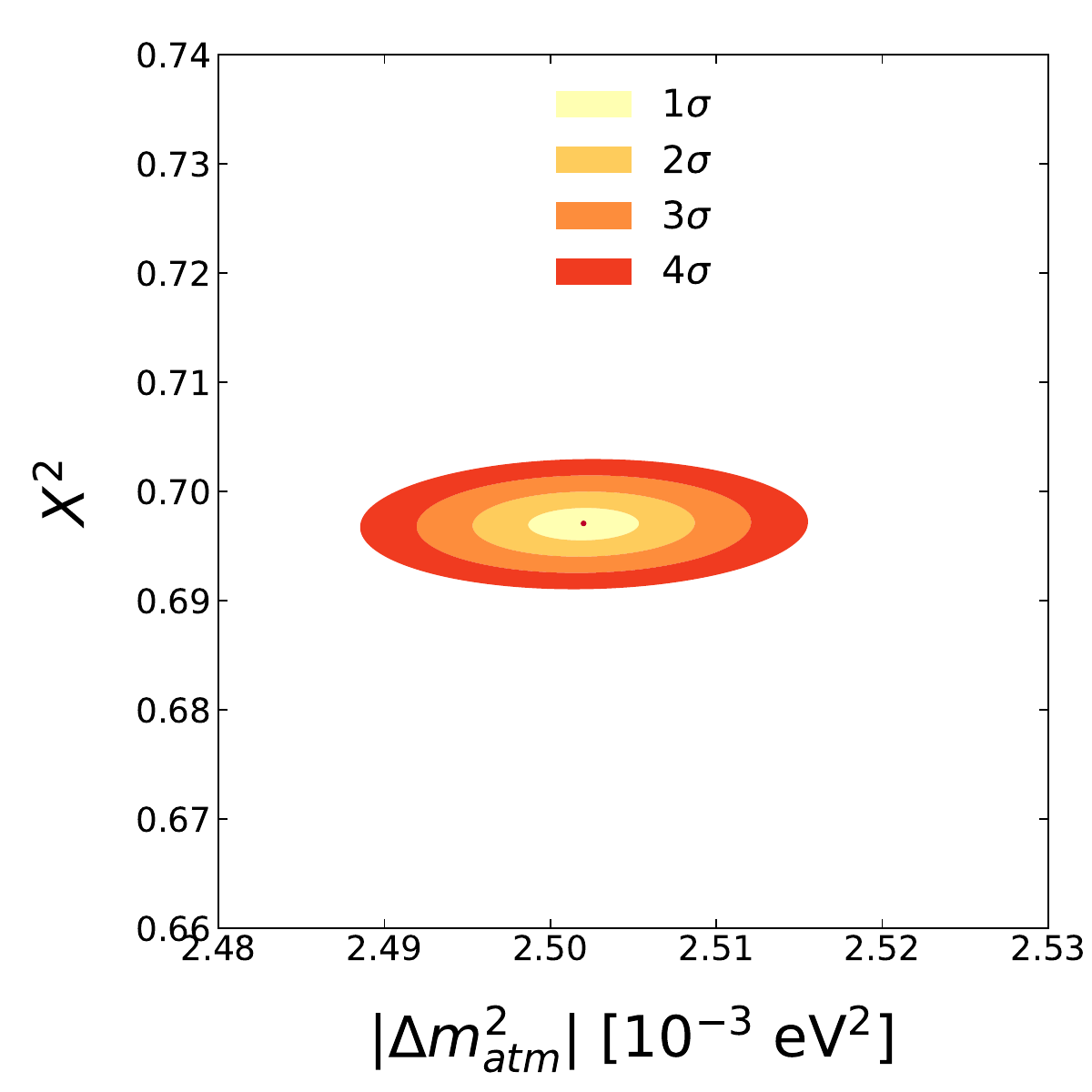}}
{\includegraphics[width=7.8cm]{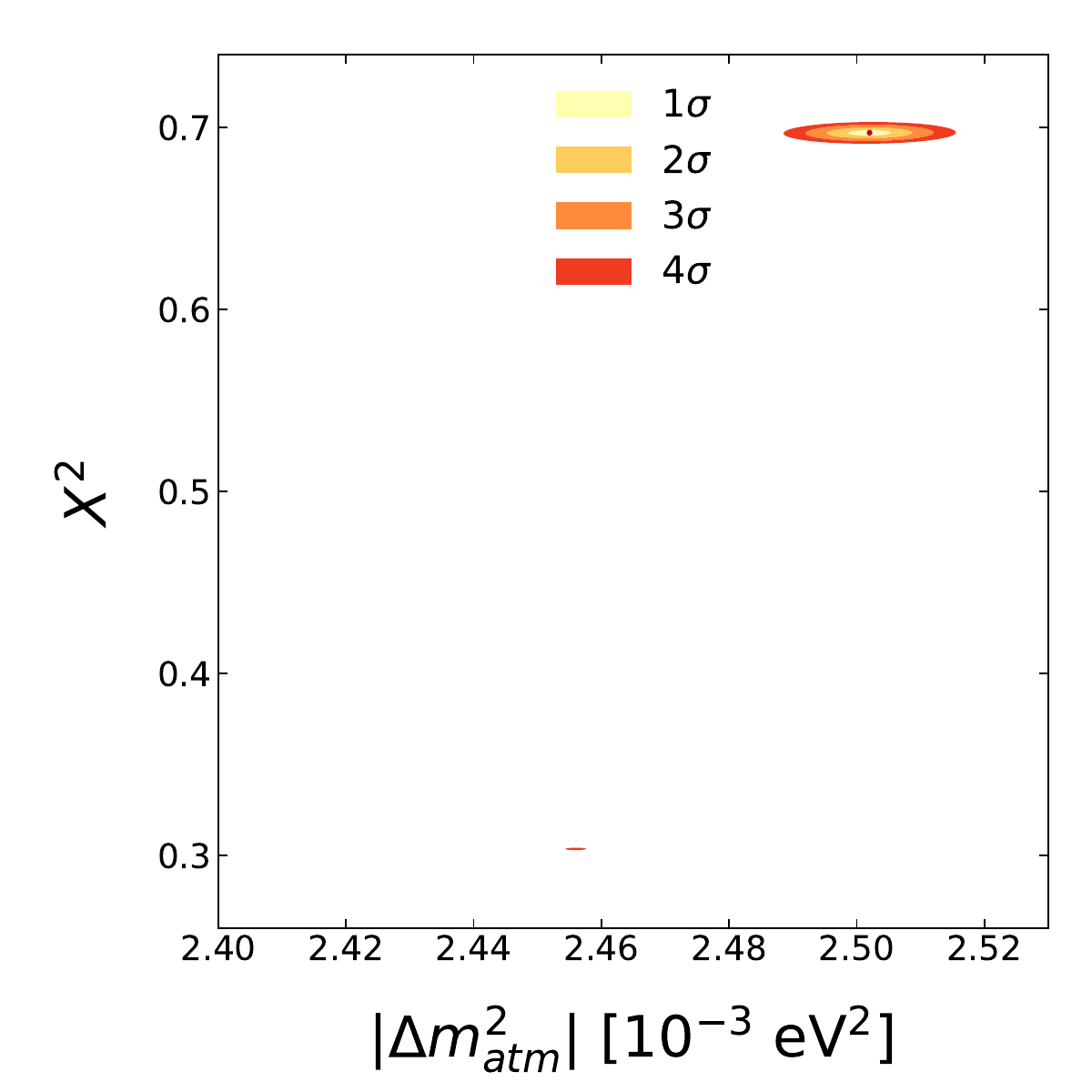}}
\end{center}
\vskip -0.4cm
\caption{
The same as in Fig. \ref{Fig1} but assuming 
IO neutrino mass spectrum to be the "true" one.  
The second minimum barely seen in 
the right panel at 4$\sigma$ C.L.  at $X^2 = 0.303$ 
corresponds to NO neutrino mass spectrum. 
See text for further details.
}
\label{Fig2}
\vspace{-0.1cm}
\end{figure}
%
\noindent
be used as input in the proposed alternative 
analysis of the relevant (prospective) 
reactor neutrino data.
However, $\Delta{m}^2_{\mbox{atm}}$ has to be determined 
together with the parameter $X^2$ 
in the corresponding statistical analysis.

We have performed a statistical analysis 
to evaluate the potential sensitivity 
of the proposed alternative method of 
neutrino mass ordering determination. 
The analysis 
has the following characteristics.
The ``true" spectrum of events $S^*$($E_{\text{vis}}$) is calculated 
for the best fit values of the oscillation parameters
given in eqs. (\ref{deltasolvalues}) - (\ref{theta13values})
and for either normal or inverted ordering, i.e., 
for $X^2=(\sin^2\theta_{12})_{\text{BF}}$ or 
$X^2=1-(\sin^2\theta_{12})_{\text{BF}}$. 
The statistical component of the $\chi^2$ is obtained comparing 
the "true" spectrum $S^*$
 with a family of spectra $S$($E_{\text{vis}}$) obtained by varying 
the parameters ($\Delta m^2_\odot$, 
 $\Delta m^2_{\text{atm}}$, $\theta_{13}$, $X^2$), through the equation
\begin{equation}
 \chi^2_{\text{stat}}=\int dE_{\text{vis}}\left(\frac{S^*(E_{\text{vis}})-S(E_{\text{vis}})}{\sqrt{S^*(E_{\text{vis}})}}\right)^2\ ,
 \label{stat_chi2}
 \end{equation}
%
where we are taking the limit 
of an infinite number of bins, which is 
practically  valid 
numerically with $\gtrsim$ 250 bins.
As a case study we 
consider the JUNO experiment \cite{JUNO}, 
with a total exposition of 
3.6$\times 10^3$ GW$\times$kton$\times$yr 
and an energy resolution
\begin{equation}
\frac{\sigma_e(E_e)}{E_e+m_e} = 
\frac{2.57\times10^{-2}}{\sqrt{(E_e+m_e)/\text{MeV}}}+0.18\times 10^{-2}\ .
\label{energy_resolution}
\end{equation}
%
The calculation of $S$ and $S^*$ follows the approach employed in 
\cite{Capozzi:2013psa,Capozzi:2015bpa} including a background from 
geo-neutrinos and two far reactors.
For simplicity we assumed neutrino oscillations in vacuum and 
the presence of only one nuclear reactor contributing to 
the respective event rate.
As systematic uncertainties we 
take into account three normalisation errors: 
one regarding the reactor flux uncertainty (3\%), 
and two related to the normalization of the Uranium (20\%) and 
Thorium (27\%) components of the geo-neutrino flux, respectively.
We are not considering energy scale \cite{Ghoshal:2010wt} and 
flux shape uncertainties, 
as done, e.g., in \cite{Capozzi:2015bpa}, 
which is beyond the scope of the present study. 
We also include uncertainties on $\Delta{m}^2_{\odot}$, $\sin^2\theta_{13}$ 
and $\Delta{m}^2_{\text{atm}}$, according to the 
1$\sigma$ errors reported in eqs. (\ref{deltasolvalues}) - 
(\ref{theta13values}). To each of the above systematic uncertainties
there corresponds a quadratic penalty 
$(p-p^*)^2/\sigma_p^2$, where $p$ represents
a generic systematic parameter, $p^*$ is its ``true" value and $\sigma_p$
is its 1$\sigma$ error. The quantity $X^2$ is considered a 
free parameter, i.e., without external constraints.
The final $\chi^2$ corresponding to a given point of the
parameter space is then obtained  by 
$\chi^2=\chi^2_{\text{stat}}+\chi^2_{\text{par}}$,
where the second contribution is the sum of the quadratic penalties.

  In Fig. \ref{Fig1}, left panel, we  show the 1$\sigma$, 2$\sigma$, 
3$\sigma$ and 4$\sigma$ allowed regions (for 1 degree of freedom) in the plane
$\Delta m^2_{\text{atm}}$--$X^2$ assuming normal ordering. The red point
represents the best fit point, which corresponds by construction to the values
of $\sin^2\theta_{12}$ and $\Delta m^2_{\text{atm}}$
reported in eqs. (\ref{sinsolvalues}) and (\ref{deltaatmvalues}). 
In Fig. \ref{Fig1}, right panel, the presence of a second 
local $\chi^2$ minimum at  
$X^2 = 1-(\sin^2\theta_{12})_{\text{BF}} = 0.695$ 
(and $|\Delta m^2_{\text{atm}}| = 2.565\times 10^{-3}$ eV$^2$)
associated with the spectrum with inverted ordering, 
is also seen. This minimum, as is indicated by Fig.  \ref{Fig1}, 
appears at approximately 4$\sigma$ C.L. More precisely, 
we find for the difference between the $\chi^2$ minima 
in the NO and IO cases
$\chi^2_{\rm min}(IO) - \chi^2_{\rm min}(NO) = 14.5$. 
This implies that the type of neutrino mass ordering can be 
established at 3.8$\sigma$ C.L.

 Performing a similar analysis but assuming the 
inverted ordering to be the "true" one, 
the results of which are shown graphically 
on Fig. \ref{Fig2},  we find 
$\chi^2_{\rm min}(NO) - \chi^2_{\rm min}(IO) = 15.8$, 
i.e., that the type of neutrino mass ordering 
can be established at $3.975\sigma \cong 4.0\sigma$ C.L.
   
These confidence levels are somewhat higher than the 
3$\sigma$ C.L. which the analyses using the standard 
method (and performed under the same conditions)  
typically give (see, e.g., \cite{Capozzi:2020cxm}).
The reason the alternative method employed 
in our analysis leads to a somewhat 
stronger rejection of the "wrong" ordering 
can be related to the fact that 
i) we have neglected matter effects in the 
oscillations of reactor $\bar{\nu}_e$, which, 
although small, have to be taken into account 
given the exceptionally high precision of 
JUNO experiment, and 
ii) we have not taken into account the small difference 
between the baselines of the reactors which provide the 
flux of  $\bar{\nu}_e$ for JUNO. 

 Our results show that the sensitivity to 
the neutrino mass ordering that can be achieved employing 
the proposed alternative method of determination of the ordering 
in any case is not worse than the sensitivity that can achieved using the 
standard approach \cite{Petcov:2001sy}.
The precision with which $\sin^2\theta_{12}$ can be determined 
using the proposed alternative approach, as 
Figs. \ref{Fig1} and  \ref{Fig2} indicate, 
is exceptionally high and matches the precision that  
can be reached utilising the standard method \cite{Choubey:2003qx}: 
we find that  $\sin^2\theta_{12}$ can be determined with 
1$\sigma$ relative uncertainty of 0.49\% 
in both cases of true spectrum considered
\footnote{In the version of our article submitted to the archive in 2017  
(see arXiv:1701.06328v1)
and published in Phys. Lett. B772 (2017) 179,
there was an error in incorporating the expression of the probability 
given in eq. (\ref{PNOIO}) in the code used for the statistical analysis.
As a consequence, Fig. 1 and the sensitivity on $\sin^2\theta_{12}$ 
reported in arXiv:1701.06328v1 and in the published version 
of the article are incorrect. They are corrected here using 
updated values of the neutrino oscillation parameters 
quoted in eqs. (\ref{deltasolvalues}) - (\ref{theta13values}). 
We have added also Fig. 2 for completeness of the analysis. 
We would like to thank Yuanyuan Zhang for useful correspondence 
regarding the problematic Fig. 1 in arXiv:1701.06328v1.
}.
The precision on  $\Delta{m}^2_{\text{atm}}$, 
which is determined simultaneously with 
$\sin^2\theta_{12}$ (or $X^2$), 
as Figs. \ref{Fig1} and \ref{Fig2} show, is also exceptionally high: 
assuming NO (IO) spectrum to be the "true" one 
we get for the  1$\sigma$ relative uncertainty 
0.15\% (0.14\%). 
This can be used, in addition to the 
determination of the value of 
$\sin^2\theta_{12}$ (or $X^2$), for 
establishing the correct neutrino mass ordering 
by comparing the remarkably precise 
value of  $\Delta{m}^2_{\text{atm}}$ 
obtained using the alternative 
method with the values of $\Delta{m}^2_{\text{atm}}$ 
determined in the cases of NO and IO spectra in the 
long baseline accelerator and atmospheric 
neutrino oscillation experiments.

We have found out further 
that the requisite detector energy resolution 
for successful determination 
of the neutrino mass ordering at least at the 
3$\sigma$ C.L. using the discussed 
alternative method is the same as 
that required when employing the standard 
approach \cite{Petcov:2001sy,Choubey:2003qx,Ghoshal:2010wt}.
These results suggest that 
the considered novel method of determination of the 
neutrino mass ordering (spectrum) can be used 
as a complementary method of the ordering (spectrum) 
determination independently of, 
and on a equal footing with, 
the standard method.

\section{Summary} 

 In the present article we have investigated an alternative 
method of determination the type of spectrum neutrino masses obey 
in medium baseline experiments with reactor neutrinos.  
The study was performed within the reference 3-neutrino mixing scheme.
Within this scheme  the neutrino mass spectrum, as is well known,  
can be of two varieties: with normal ordering (NO) and 
with inverted ordering (IO). The two possible types 
of neutrino mass spectrum are related to the two possible signs 
of the atmospheric neutrino mass squared difference 
$\Delta m^2_{31(32)}$, allowed by the existing global 
neutrino oscillation data.
The new method is based on the observation that 
the expressions for the relevant $\bar{\nu}_e$ 
survival probability in the cases of NO and IO 
spectra,   $P^{NO}(\bar{\nu}_e \rightarrow \bar{\nu}_e)$ 
and $P^{IO}(\bar{\nu}_e \rightarrow \bar{\nu}_e)$, 
differ only by one factor $X^2$ 
in the interference term involving 
the solar neutrino  and the atmospheric neutrino 
mass squared differences, $\Delta m^2_{\odot}(=\Delta m^2_{21})$ 
and $\Delta m^2_{\mbox{atm}}$ 
(see eqs. (\ref{PNO}) - (\ref{PNOIO})):
$X^2 = \sin^2\theta_{12}$ in the NO case and 
$X^2 = \cos^2\theta_{12}$ in the IO one, 
$\theta_{12}$ being the solar neutrino mixing angle. 
In the standard approach of determining the neutrino mass 
ordering (spectrum), one is supposed to use the data on 
the relevant neutrino oscillation parameters,
$\sin^{2}\theta_{12}$, $\sin^2\theta_{13}$, 
$\Delta{m}^2_{\mbox{atm}}$ and $\Delta{m}^2_{\odot}$, 
including the errors with which they are determined,
and the two different expressions for the $\bar{\nu}_e$ 
survival probability, 
$P^{NO}(\bar{\nu}_e \rightarrow \bar{\nu}_e)$ 
and $P^{IO}(\bar{\nu}_e \rightarrow \bar{\nu}_e)$, 
in the analysis of the data of 
the corresponding experiment (JUNO) 
with reactor $\bar{\nu}_e$.
The alternative method discussed in the present article 
consists instead of treating $X^2$ as a free parameter,  
to be determined in the data analysis 
and the result compared with the value of 
$\sin^2\theta_{12}$ found in the solar neutrino and 
KamLAND experiments. Since, according to the current data, 
the best fit values of 
$\sin^2\theta_{12}$ and  $\cos^2\theta_{12}$
differ by a factor of 2.3, 
and $\sin^2\theta_{12}$ is determined with 
a $1\sigma$ uncertainty of approximately 4.5\%, 
the proposed new method of establishing the neutrino mass 
ordering appears to be feasible. Moreover,
since $X^2$ is a continuous parameter, 
one can use standard statistical methods of 
extracting the value of $X^2$ and its respective 
uncertainty from the data.

 To test the feasibility of the proposed  
alternative method of neutrino mass ordering (spectrum) 
determination we have performed a 
statistical analysis of prospective 
(simulated) reactor neutrino data obtained with JUNO
detector. The prospective data was generated 
assuming NO (IO) neutrino mass 
spectrum and statistics corresponding 
to 3.6$\times 10^3$ GW$\times$kton$\times$yr. 
In the analysis both $X^2$ and $\Delta m^2_{\text{atm}}$ 
were determined using the simulated data. 
The results of our analysis show that 
with the "data" used  as input  
and depending on the assumed 
"true" ordering (spectrum) -- NO or IO --
the type of neutrino mass ordering (spectrum) 
can be established at the $3.8\sigma$ C.L. 
or $3.975\sigma \cong 4.0\sigma$ C.L., respectively 
(see Figs. \ref{Fig1} and \ref{Fig2}).
These confidence levels are somewhat higher than the 
3$\sigma$ C.L. which the analyses using the standard 
method (and performed under the same conditions)  
typically give (see, e.g., \cite{Capozzi:2020cxm}) 
and we discussed the possible origins of this difference.
Our results show that the sensitivity to 
the neutrino mass ordering that can be achieved employing 
the proposed alternative method of determination of the ordering 
in any case is not worse than the sensitivity that can achieved 
using the standard approach \cite{Petcov:2001sy}. 
The precision with which 
$\sin^2\theta_{12}$ can be determined using  
the alternative method considered in the present article  
(as Figs. \ref{Fig1} and \ref{Fig2} indicate) is 
exceptionally high and matches  
the precision on $\sin^2\theta_{12}$ that can be reached 
utilising the standard  method
(see, e.g., \cite{Choubey:2003qx}):
we find that $\sin^2\theta_{12}$ can be  determined with 
1$\sigma$ relative uncertainty of 0.49\% 
in both cases of true spectrum considered.

The precision on  $\Delta{m}^2_{\text{atm}}$, 
which is determined simultaneously with 
$\sin^2\theta_{12}$ (or $X^2$), 
as Figs. \ref{Fig1} and \ref{Fig2} show, is also exceptionally high: 
assuming NO (IO) spectrum to be the true one 
we get for the  1$\sigma$ relative uncertainty 
0.15\% (0.14\%). This can be used, in addition to the 
determination of the value of $\sin^2\theta_{12}$ 
(or $X^2$), for establishing the correct neutrino mass ordering 
by comparing the value of  $\Delta{m}^2_{\text{atm}}$,  
obtained potentially with remarkable precision 
using the alternative method, with the values 
of $\Delta{m}^2_{\text{atm}}$ found in the NO and IO cases in 
the long baseline accelerator and atmospheric 
neutrino oscillation experiments.
The analyses performed by us show also that 
the requisite detector energy resolution 
for successful determination 
of the neutrino mass ordering at least at the 
3$\sigma$ C.L. using the discussed 
alternative method is approximately the same as 
that required when employing the standard 
approach, i.e., it should be not worse than 
approximately $3\%/\sqrt{E_{\rm vis}/\text{MeV}}$.

 The results obtained in the present study 
suggest that the discussed alternative method of 
determination of the neutrino mass ordering (spectrum) 
in medium baseline reactor neutrino experiments can be used as 
an additional independent method of neutrino mass 
ordering determination on an equal footing with and complementary to 
the standard one discussed in 
\cite{Petcov:2001sy,Choubey:2003qx,Learned:2006wy,Zhan:2008id,Ghoshal:2010wt,Qian:2012xh,Ghoshal:2012ju,Ciuffoli:2012iz,Ge:2012wj,Li:2013zyd,Capozzi:2013psa,Capozzi:2015bpa,Wang:2016vua,Blennow:2013oma,Capozzi:2020cxm}.

\vspace{0.4cm}
{\bf Acknowledgements.} 
The authors would like to thank 
 L. Stanco for useful comments.
This work was supported in part by the INFN
program on Theoretical Astroparticle Physics (TASP), by the research
grant  2012CPPYP7
under the program  PRIN 2012 funded by the Italian 
Ministry of Education, University and Research (MIUR),
by the European Union Horizon 2020 research and innovation programme
under the  Marie Sklodowska-Curie grants 674896 and 690575, and by
the World Premier International Research Center Initiative (WPI
Initiative), MEXT, Japan (S.T.P.). 
S.M.B. acknowledges the support of the RFFI grant 16-02-01104.


\begin{thebibliography}{99}

\bibitem{Petcov:2001sy}
  S.~T.~Petcov and M.~Piai,
  Phys.\ Lett.\ B {\bf 533} (2002) 94
  [hep-ph/0112074].
%
\bibitem{Bilenky:2016nff}
  S.~M.~Bilenky,
  arXiv:1603.05808.

\bibitem{PDG2016} K. Nakamura and S.T. Petcov in 
M. Tanabashi  {\it et al.} [Particle Data Group Collab.],
  "Review of Particle Physics",
Phys. Rev. D {\bf 98} (2018) 030001.
%
\bibitem{NOvA2016} P.~Adamson {\it et al.} [NOvA Collaboration],
Phys. Rev. Lett. {\bf 116} (2016) 151806
[arXiv:1601.05022]; 
A. Himmel [NOvA Collaboration], 
talk at "The XXIX International Conference on Neutrino 
Physics and Astrophysics" (Neutrino 2020), 
Chicago, June 22 - July 2, 2020 (virtual conference),
DOI:5281/zenodo.3959581.
%
\bibitem{DUNE2016} 
R. Acciarri {\it et al.} [DUNE Collaboration],
arXiv:1601.05471 and arXiv:1601.02984;
M. Mooney [DUNE Collaboration], 
talk at "The XXIX International Conference on Neutrino 
Physics and Astrophysics" (Neutrino 2020), 
Chicago, June 22 - July 2, 2020 (virtual conference),
DOI:5281/zenodo.3959638.
%
\bibitem{T2HKK2016} K. Abe {\it et al.} [Hyper-Kamiokande Proto-Collab.], 
arXiv:1611.06118; Prog. Theor. Exp. Phys. {\bf 5} (2015) 053C02
[arXiv:1502.05199];
M. Ishitsuka, talk at "The XXIX International Conference on Neutrino 
Physics and Astrophysics" (Neutrino 2020), 
Chicago, June 22 - July 2, 2020 (virtual conference),
DOI:5281/zenodo.3959585.
%
\bibitem{Pascoli:2002xyz} S. Pascoli and S.T. Petcov,
Phys.\ Lett.\ B {\bf 544} (2002) 239 
[hep-ph/0205022].
%
\bibitem{JUNO} F. An {\it et al.} [JUNO Collab.],
J. Phys. G {\bf 43} (2016) 3 [arXiv:1507.05613];
Y. Meng, talk at "The XXIX International Conference on Neutrino 
Physics and Astrophysics" (Neutrino 2020), 
Chicago, June 22 - July 2, 2020 (virtual conference),
DOI:5281/zenodo.3959622.
%
\bibitem{RENO50} S.B. Kim, Nucl. Part. Phys. Proc. {\bf 256-266} (2015) 93 
[arXiv:1412.2199].
%
\bibitem{PINGU} 
M. G. Aartsen {\it et al.} [IceCube-PINGU Collab.],
arXiv:1401.2046.
%
\bibitem{ORCA2016}
  S.~Adrian-Martinez {\it et al.} [KM3Net Collab.],
  J. Phys. G {\bf 43} (2016)  084001
  [arXiv:1601.07459]. 
%
\bibitem{INO} S. Ahmed {\it et al.} [INO Collab.],
Pramana {\bf 88} (2017) 79
[arXiv:1505.07380].
%
\bibitem{BPont57} B. Pontecorvo, 
                  Zh. Eksp. Teor. Fiz. 
{\bf 33} (1957) 549 and {\bf 34} (1958) 247.
%
\bibitem{MNS62} Z. Maki, M. Nakagawa and S. Sakata, 
Prog. Theor. Phys. {\bf 28} (1962) 870.
%
\bibitem{BPont67}
  B.~Pontecorvo,
  Zh. Eksp. Teor. Fiz.  {\bf 53} (1967) 1717.
%
\bibitem{Capozzi:Nu2016} F. Capozzi {\it et al.}, 
J. Phys. Conf. Ser. {\bf 718} (2016) 062042;
Prog. Part. Nucl. Phys. {\bf 102} (2018) 48
[arXiv:1804.09678].
%
\bibitem{Esteban:2016qun}
  I.~Esteban {\it et al.},
JHEP {\bf 1701} (2017) 087
[arXiv:1611.01514];
  JHEP {\bf 1901} (2019) 106
  [arXiv:1811.05487].
%
\bibitem{Capozzi:2020qhw}
F.~Capozzi {\it et al.},
  Phys. Rev. D {\bf 101} (2020) 116013
  [arXiv:2003.08511].

\bibitem{Bilenky:2001jq}
  S.M. Bilenky, D. Nicolo and S.T. Petcov,
  Phys. Lett. B {\bf 538} (2002) 77
  [hep-ph/0112216].
%
\bibitem{Apollonio:1999ae}
  M.~Apollonio {\it et al.} [CHOOZ Collaboration],
  Phys. Lett. B {\bf 466} (1999) 415
  [hep-ex/9907037].
%
\bibitem{Choubey:2003qx}
  S.~Choubey, S.~T.~Petcov and M.~Piai,
  Phys. Rev. D {\bf 68} (2003) 113006
  [hep-ph/0306017].
%
\bibitem{Learned:2006wy}
  J.~Learned  {\it et al.},  
  Phys. Rev. D {\bf 78} (2008) 071302 
[hep-ex/0612022].
%
%

%
\bibitem{Zhan:2008id}
  L.~Zhan {\it et al.},
  Phys. Rev. D {\bf 78} (2008) 111103
  [arXiv:0807.3203],
and 
%
  Phys. Rev. D {\bf 79} (2009) 073007
  [arXiv:0901.2976].
%
\bibitem{Ghoshal:2010wt}
  P. Ghoshal and S.T. Petcov,
  JHEP {\bf 1103} (2011) 058
  [arXiv:1011.1646].
%
\bibitem{Qian:2012xh}
  X. Qian {\it et al.},
  Phys. Rev. D {\bf 87} (2013)  033005
  [arXiv:1208.1551].
%
\bibitem{Ghoshal:2012ju}
  P. Ghoshal and S.T. Petcov,
  JHEP {\bf 1209} (2012) 115
  [arXiv:1208.6473].
%
\bibitem{Ciuffoli:2012iz}
  E. Ciuffoli, J. Evslin and X. Zhang,
  JHEP {\bf 1303} (2013) 016
  [arXiv:1208.1991];
  Phys. Rev. D {\bf 88} (2013)  033017
  [arXiv:1302.0624];
%
  JHEP {\bf 1401} (2014) 095
  [arXiv:1305.5150].
%
\bibitem{Ge:2012wj}
  S. F. Ge {\it et al.},
  JHEP {\bf 1305} (2013) 131
  [arXiv:1210.8141].
%
\bibitem{Li:2013zyd}
  Y.F. Li {\it et al.},
  Phys. Rev. D {\bf 88} (2013) 013008
  [arXiv:1303.6733].
%
\bibitem{Capozzi:2013psa}
  F.~Capozzi, E.~Lisi and A.~Marrone,
  Phys. Rev. D {\bf 89} (2014)  013001
  [arXiv:1309.1638].

\bibitem{Capozzi:2015bpa}
  F.~Capozzi, E.~Lisi and A.~Marrone,
  Phys. Rev. D {\bf 92} (2015)  093011.
%
 \bibitem{Wang:2016vua}
H.~Wang {\it et al.},
Nucl. Phys. B {\bf 918} (2017) 245
[arXiv:1602.04442].
%
\bibitem{Blennow:2013oma}
  M.~Blennow {\ it et al.},
  JHEP {\bf 1403} (2014) 028
  [arXiv:1311.1822];
  M.~Blennow,
  JHEP {\bf 1401} (2014) 139
  [arXiv:1311.3183].
%
\bibitem{DBay2012}  F.P. An {\it et al.} [Daya Bay Collab.],
Phys. Rev. Lett. {\bf 108} (2012) 171803.
%
\bibitem{RENO2012}  J.K. Ahn {\it et al.} [RENO Collab.], 
Phys. Rev. Lett. {\bf 108} (2012) 191802.
%
\bibitem{abe11b} K. Abe {\it et al.} [T2K Collab.], 
Phys. Rev. Lett. {\bf 107} (2011) 041801
[arXiv:1106.2822].
%
\bibitem{adamson11c} P. Adamson {\it et al.} [MINOS Collab.],
Phys. Rev. Lett. {\bf 107} (2011) 181802 
[arXiv:1108.0015].
%
\bibitem{DChoozth13} Y. Abe {\it et al.} [Double Chooz Collab.],
Phys. Rev. Lett. {\bf 108} (2012) 131801
[arXiv:1112.6353].
%
\bibitem{Capozzi:2020cxm}
  F.~Capozzi, E.~Lisi and A.~Marrone,
  arXiv:2006.01648.



\end{thebibliography}
\end{document}